\documentclass[preprintnumbers,amsmath,amssymb,showpacs]{revtex4}

\usepackage{graphicx}
\usepackage{dcolumn}
\usepackage{bm}

\begin{document}
\title{A measure of multipartite entanglement with computable lower bounds}

\author{Yan Hong}
\author{Ting Gao}
\email{gaoting@hebtu.edu.cn}
\affiliation {College of Mathematics and Information Science, Hebei
Normal University, Shijiazhuang 050024, China}
\author{Fengli Yan}
\email{flyan@hebtu.edu.cn}
\affiliation {College of Physics Science and Information Engineering, Hebei
Normal University, Shijiazhuang 050024, China}

\date{\today}

\begin{abstract}
In this paper, we present a measure of multipartite entanglement ($k$-nonseparable), $k$-ME concurrence $C_{k-\mathrm{ME}}(\rho)$ that unambiguously detects all $k$-nonseparable states in arbitrary dimensions,  where the special case, 2-ME concurrence $C_{2-\mathrm{ME}}(\rho)$, is a measure of genuine multipartite entanglement. The new measure $k$-ME concurrence  satisfies important characteristics of an entanglement measure including entanglement monotone, vanishing on $k$-separable states, convexity, subadditivity and strictly greater than zero for all $k$-nonseparable states.
 Two powerful lower bounds on this measure are given. These lower bounds are experimentally implementable
without quantum state tomography and are easily computable as no optimization or eigenvalue evaluation is
needed. We illustrate detailed examples in which the given bounds perform better than other known detection criteria.
\end{abstract}


\maketitle

\section{Introduction}
Entanglement as a physical resource plays an important role in quantum information, such as, quantum communication \cite{Ekert91, BBCJPWteleportation93, BBM92, LoChau, YZEPJB05, GYJPA05, WangPhysRep, YGPRA11,GaoYanEPL08} and quantum computing \cite{RB,BD}.
So it is a significant work to quantify entanglement not only in  theoretical research but also in practical application. One of the main goals of the theory of entanglement is to develop measures of entanglement.
Several entanglement measures \cite{HorodeckisRMP2009, GuhneToth2009, PlenioVirmaniQIC07}  have been introduced, such as entanglement distillation \cite{BennettBernstein96, BennettBrassard96, BennettDiVincenzo96}, entanglement cost \cite{BennettDiVincenzo96, HaydenHorodecki2001}, entanglement of formation \cite{BennettDiVincenzo96,Shor2003}, negativity \cite{ZHSLpra1998, VWpra2002}, three-tangle \cite{CKW2000} and localizable entanglement \cite{VerstraetePopCiracPRL04, GaoYanEPL08}. These measures except localizable entanglement are entanglement monotones \cite{HorodeckisRMP2009, GuhneToth2009, PlenioVirmaniQIC07}, in that they cannot increase under local operations and classical communication (LOCC), whereas localizable entanglement can deterministically increase under LOCC operations between all parties \cite{GSpra06}. In bipartite setting, entanglement cost, entanglement of formation, and negativity are convex, moreover, entanglement cost, entanglement of formation are also subadditive. It is an open question
whether entanglement distillation is convex \cite{HorodeckisRMP2009}. The negativity fails to recognize entanglement in PPT states. In the multipartite setting, three-tangle is invariant under permutation of the three systems and is in fact an entanglement monotone for three-qubit systems. However, there are states with genuine three party entanglement for which the three-tangle can be zero (the W-state serves as an example \cite{CKW2000}), i.e., the three-tangle has the disadvantageous property that it vanishes for some entangled states. Localizable entanglement \cite{VerstraetePopCiracPRL04} requires an underlying measure of bipartite entanglement to quantify the entanglement between the two singled-out parties. When concurrence was used as underlying measure of bipartite entanglement, Gao \textit{et al.} \cite{GaoYanEPL08} derived an easily computable formula for localizable entanglement in the three-qubit case.

 The concurrence  is a very popular measure for the quantification of bipartite quantum correlations \cite{HWprl97, Wootters1998, GuhneToth2009,HorodeckisRMP2009}, and is also defined for bipartite high dimensional states \cite{MintertKus2005}, but it is not computable because of optimization for bipartite high dimensional mixed states.
For multipartite quantum systems, although there are some criteria \cite{GuhneToth2009, KrammerKampermann2009, Seevinck2010, HuberMintert2010, GaoHong2010, GaoHong2011, GaoHong2012, HiesmayrHuber2010, HuberSchimpf2011, HuberErker2011,GHGcovariancematrix10} to detect genuine multipartite entanglement, but there is not computable measure quantifying the amount of  multipartite entanglement in general. Ma  $et~al.$ \cite{MaChen2011} defined a generalized concurrence called GME-concurrence which satisfies the necessary conditions for genuine multipartite
entanglement measure \cite{MintertCarvalho2005, HiesmayrHuber2009}. Although for general mixed states it is not computable owing to the optimization, they gave lower bounds \cite{MaChen2011, ChenMa2012}. What we are looking for is multipartite entanglement measure such that its values vanish with respect to $k$-separable states, whereas they are strictly positive for $k$-nonseparable states.

In this paper, we introduce a generalized concurrence ($k$-ME concurrence) for a finite-dimensional systems of arbitrarily many parties as an entanglement measure, which satisfies important characteristics of an entanglement measure, such as entanglement monotone, vanishing on $k$-separable states, invariant under local unitary
transformations, convexity, subadditivity and strictly greater than zero for all $k$-nonseparable states. This multipartite entanglement measure unambiguously detects all $k$-nonseparable states in arbitrary dimensions.
The GME concurrence  \cite{ChenMa2012,MaChen2011} is the special case of our $k$-ME concurrence when $k=2$.   We show that strong lower bounds on this measure can be derived by exploiting close analytic relations between this concurrence and recently introduced detection criteria for multipartite entanglement \cite{GaoHong2010, GaoHong2011, GaoHong2012}. And then we provide examples in which the entanglement criteria based on our lower bound have better performance with respect to the known methods, the lower bounds obtained by Refs. \cite{MaChen2011, ChenMa2012}.

\section{Multipartite entanglement}

Before we state the definition of multipartite entanglement measure, $k$-ME concurrence, and its lower bounds, an introduction of
concepts and notations that will be involved in the subsequent
sections of our article is necessary. Throughout the paper, we consider a multiparticle quantum system $\mathcal{H}=\otimes_{i=1}^n\mathcal{H}_i=\mathcal{H}_{1}\otimes\mathcal{H}_{2}\otimes\cdots\otimes\mathcal{H}_{n}$  with $n$ parts of respective dimension $d_i$, $i=1, 2, \cdots, n$. A $k$-partition $A_1|A_2|\cdots|A_k$ (of $\{1,2,\cdots, n\}$) means that the set $\{A_1, A_2, \cdots, A_k\}$ is a collection of pairwise disjoint sets, and the union of all sets in $\{A_1, A_2, \cdots, A_k\}$ is $\{1,2,\cdots, n\}$ (disjoint union $\bigcup\limits_{i=1}^kA_i=\{1,2,\cdots, n\}$).
An  pure state $|\psi\rangle$ of an $n$-partite quantum system $\mathcal{H}$ is called $k$-separable if there is a $k$-partition $A_1|A_2|\cdots|A_k=j_1^1\cdots j_{n_1}^1|j_1^2\cdots j^2_{n_2}|\cdots|j_1^k\cdots j^k_{n_k}$ such that
\begin{equation}\label{k-separablePureState}
|\psi\rangle=|\psi_1\rangle_{A_1}|\psi_2\rangle_{A_2}\cdots|\psi_k\rangle_{A_k},
\end{equation}
where $|\psi_i\rangle_{A_i}$ is the state of subsystem $A_i$, and disjoint union $\bigcup\limits^k_{t=1}A_t =\bigcup
\limits_{t=1}^k\{j_1^t,j_2^t,\cdots,j_{n_t}^t\}=\{1,2,\cdots, n\}$.
An $n$-partite mixed state $\rho$ is $k$-separable if it can be
written as a convex combination of $k$-separable pure states
\begin{equation}\label{}
 \rho=\sum\limits_{m}p_m|\psi_m\rangle \langle\psi_m|,
\end{equation}
where $\{|\psi_m\rangle\}$ might be $k$-separable with respect to different
partitions. Thus, a mixed $k$-separable state does not need to be separable under any particular $k$-partition. In general, $k$-separable mixed states are not separable with regard to any
specific partition. If an $n$-partite state is not 2-separable (biseparable), then it is  called genuinely
$n$-partite entangled. It is called  fully
separable, iff it is $n$-separable.

Note that whenever a
state is $k$-separable, it is automatically also $k'$-separable for all $1<k'<k$. If we denote the set of all $k$-separable states by $S_k$ ($k=2,3,\cdots,n$) and
the set of all states by $S_1$, then each set $S_k$ is convex and embedded within the next set: $S_n\subset S_{n-1}\subset\cdots \subset S_2\subset S_1$,  and the complement $S_1\setminus S_k$ of $S_k$ in $S_1$ is the set of all $k$-nonseparable states. In particular, the complement $S_1\setminus S_2$ is the set of all genuine $n$-partite entangled (2-nonseparable) states. We can illustrate the convex nested
structure of multipartite entanglement in Fig. 1.

\begin{figure}
\begin{center}
{\includegraphics[scale=0.36]{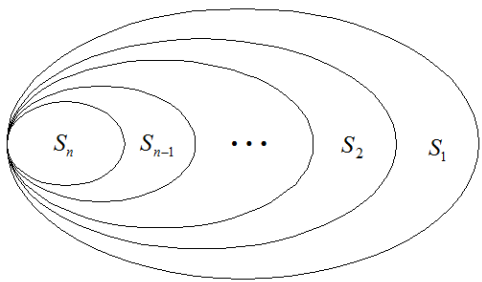}} \caption[Illustration of the convex nested structure of the sets $S_k$ of all $k$-separable states]{(Color
online). Illustration of the convex nested structure of the sets $S_k$ of all $k$-separable states. Each set is convexly embedded within the next set: $S_n\subset S_{n-1}\subset\cdots \subset S_2\subset S_1$, and the complement $S_1\setminus S_k$ of $S_k$ in $S_1$ is the set of all $k$-nonseparable states.
}
\end{center}
\end{figure}

\section{A measure of multipartite entanglement and its lower bounds}

Let us now introduce a measure of multipartite entanglement ($k$-nonseparable) that unambiguously detects all $k$-nonseparable states in arbitrary dimensions.
 For $n$-partite pure state $|\psi\rangle\in\mathcal{H}_1\otimes
\mathcal{H}_2\otimes\cdots\otimes\mathcal{H}_n$, where
dim$\mathcal{H}_l=d_l$, $l=1,2,\cdots, n$, we define the $k$-ME concurrence  as
\begin{equation}\label{}
C_{k-\mathrm{ME}}(|\psi\rangle)=\min\limits_{A}\sqrt{2\left(1-\frac{\sum\limits_{t=1}^k\textrm{Tr}(\rho^2_{A_t})}{k}\right)}
=\min\limits_{A}\sqrt{\frac{2\sum\limits_{t=1}^k\left(1-\textrm{Tr}(\rho^2_{A_t})\right)}{k}},
\end{equation}
where $\rho_{A_t}=\mathrm{Tr}_{\bar{A_t}}(|\psi\rangle\langle\psi|)$ is the reduce density matrix of subsystem $A_t$ ($\bar{A_t}$ is the complement of $A_t$ in $\{1,2,\cdots,n\}$), and the minimum is taken over all possible $k$-partitions $A={A_1}|\cdots|{A_k}$ of $\{1,2,\cdots,n\}$.    Obviously, $C_{k-\mathrm{ME}}(|\psi\rangle)$ does not only depend on $|\psi\rangle$, but also on the number $k$. However, it is independent of $k$-partitions. It should be pointed out that $C_{k-\mathrm{ME}}(|\psi\rangle)$ is non-vanishing if and only if $|\psi\rangle$ is $k$-nonseparable, that is, $C_{k-\mathrm{ME}}(|\psi\rangle)$ equals to zero if and only if $|\psi\rangle$ is $k$-separable.

For  $n$-partite mixed state $\rho$, we define the $k$-ME concurrence  as
\begin{equation}\label{kmix}
C_{k-\mathrm{ME}}(\rho)=\inf\limits_{\{ p_m,|\psi_m\rangle\}}\sum\limits_mp_mC_{k-\mathrm{ME}}(|\psi_m\rangle),
\end{equation}
where the infimum is taken over all possible pure states decompositions $\rho=\sum\limits_{m}p_m|\psi_m\rangle \langle\psi_m|$.  Specially, when $k=2$, $C_{2-\mathrm{ME}}(\rho)$ is a measure of genuine multipartite entanglement.       Note that the GME concurrence \cite{MaChen2011} is our  special case  $C_{2-\mathrm{ME}}(\rho)$, and the GME concurrence $C_{\textrm{GME}}$ is equal to $\frac{1}{\sqrt{2}}C_{2-\textrm{ME}}(\rho)$.

$k$-ME concurrence $C_{k-\textrm{ME}}(\rho)$, a measure of multipartite entanglement, satisfies the following useful properties:

M1 $C_{k-\mathrm{ME}}(\rho)=0$ for any $\rho\in S_k$ (vanishing on all $k$-separable states).

M2 $C_{k-\mathrm{ME}}(\rho)>0$ for any $\rho\in S_1\backslash S_k$ (strictly greater than zero for all $k$-nonseparable states).

M3 $C_{k-\mathrm{ME}}(U_{\textrm{Local}}^\dag\rho U_{\mathrm{Local}})=C_{k-\mathrm{ME}}(\rho)$ (invariant under local unitary
transformations).

M4 $C_{k-\mathrm{ME}}(\Lambda_{\mathrm{LOCC}}(\rho))\leq C_{k-\mathrm{ME}}(\rho)$ (entanglement monotone: nonincreasing under local operations and classical communication (LOCC)).

M5 $C_{k-\mathrm{ME}}(\sum_i p_i\rho_i)\leq\sum_ip_iC_{k-\mathrm{ME}}(\rho_i)$ (convexity).

M6  $C_{k-\mathrm{ME}}(\rho\otimes\sigma) \leq C_{k-\mathrm{ME}}(\rho)+C_{k-\mathrm{ME}}(\sigma)$  (subadditivity).

\section{Lower bounds}

\subsection{Statement of results}

Let $|\phi(x)\rangle=\otimes_{i=1}^n|x_i\rangle=|x_1x_2\cdots x_n\rangle$ be a fully separable state on Hilbert space $\mathcal{H}=\mathcal{H}_1\otimes \mathcal{H}_2\otimes\cdots\otimes\mathcal{H}_n$, and $|\Phi_{ij}(x)\rangle=|\phi_i(x)\rangle|\phi_j(x)\rangle$ a product state in $\mathcal{H}^{\otimes 2}$, where $|\phi_i(x)\rangle=|x_1x_2\cdots x_{i-1}x'_ix_{i+1}\cdots x_n\rangle$ and $|\phi_j(x)\rangle=|x_1x_2\cdots x_{j-1}x'_jx_{j+1}\cdots x_n\rangle$ are the fully separable states obtained from $|\phi(x)\rangle$ by applying (independently) local unitary transformations to $|x_i\rangle\in\mathcal{H}_i$ and $|x_j\rangle\in\mathcal{H}_j$, respectively. Let $P_{tot}$ denote the operator
that performs a simultaneous local permutation on all subsystems in
$\mathcal{H}^{\otimes 2}=(\mathcal{H}_1\otimes
\mathcal{H}_2\otimes\cdots\otimes\mathcal{H}_n)^{\otimes 2}$, while
$P_i$ just performs a permutation on $\mathcal{H}_i^{\otimes 2}$ and
leaves all other subsystems unchanged. That is, $P_{tot}=P_1\circ P_2\circ \cdots \circ P_n$, where $P_i$
 is the operator swapping the two copies of $\mathcal{H}_i$ in $\mathcal{H}^{\otimes 2}$.  For instance, $P_{tot}|x_1x_2 \cdots x_n\rangle|y_1y_2\cdots y_n\rangle=|y_1y_2\cdots y_n\rangle|x_1x_2 \cdots x_n\rangle$, while  $P_i|x_1\cdots x_{i-1}x_ix_{i+1}\cdots x_n\rangle|y_1\cdots y_{i-1}y_iy_{i+1}\cdots y_n\rangle=|x_1\cdots x_{i-1}y_ix_{i+1}\cdots x_n\rangle|y_1\cdots y_{i-1}x_iy_{i+1}\cdots y_n\rangle$.
Let
\begin{equation}\label{k-separable}
\begin{array}{rl}
  I_k(\rho,\phi(x))= & \sum\limits_{i\neq j}\sqrt{\langle\Phi_{ij}(x)|\rho^{\otimes
2}P_{tot}|\Phi_{ij}(x)\rangle}-\sum\limits_{i\neq
j}\sqrt{\langle\Phi_{ij}(x)|P_i^+\rho^{\otimes
2}P_i|\Phi_{ij}(x)\rangle} \\
  & -(n-k)\sum\limits_{i}\sqrt{\langle\Phi_{ii}(x)|P_i^+\rho^{\otimes
2}P_i|\Phi_{ii}(x)\rangle},
\end{array}
\end{equation}
then we have the following bounds.

Bound 1.

\begin{equation}\label{bound-1}
 C_{k-\mathrm{ME}}(\rho)\geq H_kI_k(\rho,\phi(x)),
\end{equation}
where \begin{equation}\label{}
H_{k}=\min\limits_{A}\frac{\sqrt{k}}{\sqrt{\sum\limits_{t=1}^kn_t(n-n_t)}}=\min\limits_{\substack{\sum\limits_{t=1}^k n_t=n}}\frac{\sqrt{k}}{\sqrt{n^2-\sum\limits_{t=1}^k n_t^2}}.
\end{equation}
Here the minimum is taken over all possible $k$-partitions $A={A_1}|\cdots|{A_k}$ of $\{1,2,\cdots,n\}$, and $n_t$ is the number of elements in $A_t$.

Specially, when $k=2$, there is
\begin{equation}\label{h2}
 H_2= \left\{ \begin{array}{ll}
                               \frac{2}{n}, & n ~ \textrm{is even}, \\
                               \frac{2}{\sqrt{n^2-1}}, & n ~ \textrm{is odd}.
                             \end{array} \right.
\end{equation}
Therefore,
\begin{equation}\label{bound1k2}
 C_{2-\mathrm{ME}}(\rho)\geq \left\{ \begin{array}{ll}
                               \frac{2}{n}I_2(\rho,\phi(x)), & n ~ \textrm{is even}, \\
                               \frac{2}{\sqrt{n^2-1}}I_2(\rho,\phi(x)), & n ~ \textrm{is odd}.
                             \end{array} \right.
\end{equation}
It is stronger than the lower bound 1 in \cite{ChenMa2012}, since $H_2$ is greater than $\frac{1}{\sqrt{2}(n-1)}$. That is, our lower bound 1 is more powerful than that in \cite{ChenMa2012}.

Bound 2.

\begin{equation}\label{bound-2}
C_{k-\mathrm{ME}}(\rho)\geq\max\limits_{\{\phi(x), \phi(y)\}}\bar{H}_k(I_k(\rho,\phi(x))+I_k(\rho,\phi(y))),
\end{equation}
where
\begin{equation}\label{}
\bar{H}_{k}=\min\limits_{A}\frac{\sqrt{k}}{\sqrt{2\sum\limits_{t=1}^kn_t(n-n_t)}}=\frac{1}{\sqrt{2}}H_k.
\end{equation}
Here $|\phi(x)\rangle=\otimes_{i=1}^n|x_i\rangle$ and $|\phi(y)\rangle=\otimes_{i=1}^n|y_i\rangle$ are orthogonal full separable states.

The proof of two lower bounds above is placed in the appendix.

\subsection{Examples}

\textit{Example 1}: Consider the $n$-qubit state family given by a mixture of the identity matrix, the $W$ state and the anti-$W$ state
\begin{equation}\label{n-qubit}
\rho_n=\frac{1-2a}{2^n}I_{2^n}+a|W_n\rangle\langle W_n|+b|\tilde{W_n}\rangle\langle \tilde{W_n}|,
\end{equation}
where $|W_n\rangle=\frac{1}{\sqrt{n}}(|00\cdots001\rangle+|00\cdots010\rangle+\cdots+|10\cdots000\rangle)$ and $|\tilde{W_n}\rangle=\frac{1}{\sqrt{n}}(|11\cdots110\rangle+|11\cdots101\rangle+\cdots+|01\cdots111\rangle)$. Let $|\phi(0)\rangle=|0\rangle^{\otimes n}$ and $|\phi(1)\rangle=|1\rangle^{\otimes n}$, then $|\phi_i(0)\rangle=|0\cdots 010 \cdots 0\rangle$ and $|\phi_i(1)\rangle=|1\cdots 101 \cdots 1\rangle$ can be obtained by applying the bit-flip operation $\sigma_x$ on the $i$-th qubit of $|\phi(0)\rangle$ and $|\phi(1)\rangle$, respectively.

When $n>3$, there are
\begin{equation}\label{}
\begin{array}{rl}
I_k(\rho_n,\phi(0))=&(k-1)a-\frac{n(2n-k-1)(1-a-b)}{2^n},
\end{array}
\end{equation}
\begin{equation}\label{}
\begin{array}{rl}
I_k(\rho_n,\phi(1))=&(k-1)b-\frac{n(2n-k-1)(1-a-b)}{2^n}.
\end{array}
\end{equation}

When $n=3$, there are
\begin{equation}\label{}
\begin{array}{rl}
I_k(\rho_3,\phi(0))=&(k-1)a-\frac{3}{4}\sqrt{\frac{(1-a-b)(3-3a+5b)}{3}}-\frac{3(3-k)(1-a-b)}{2^3},
\end{array}
\end{equation}
\begin{equation}\label{}
\begin{array}{rl}
I_k(\rho_3,\phi(1))=&(k-1)b-\frac{3}{4}\sqrt{\frac{(1-a-b)(3+5a-3b)}{3}}-\frac{3(3-k)(1-a-b)}{2^3},
\end{array}
\end{equation}

Our bound 1 Ineq.(\ref{bound-1}) is
\begin{equation}\label{eg1.bound1}
C_{k-\textrm{ME}}\geq \left\{ \begin{array}{ll}
                      \max\{H_kI_k(\rho_n,\phi(0)),H_kI_k(\rho_n,\phi(1))\},  & n>3, \\
                      \max\{H_kI_k(\rho_3,\phi(0)),H_kI_k(\rho_3,\phi(1))\}, & n=3,
                     \end{array}
 \right.
\end{equation}
where $H_k=\min\limits_{\substack{\sum\limits_{t=1}^k n_t=n}}\frac{\sqrt{k}}{\sqrt{n^2-\sum\limits_{t=1}^k n_t^2}}$.

Specially,
\begin{equation}\label{eg1.Bound1k=2}
C_{2-\textrm{ME}}\geq \left\{ \begin{array}{ll}
                      \max\{\frac{2}{n}I_k(\rho_n,\phi(0)),\frac{2}{n}I_k(\rho_n,\phi(1))\},  & n>3 ~ \textrm{and} ~ n ~ \textrm{is even}, \\
                      \max\{\frac{2}{\sqrt{n^2-1}}I_k(\rho_n,\phi(0)),\frac{2}{\sqrt{n^2-1}}I_k(\rho_n,\phi(1))\},  & n>3 ~ \textrm{and} ~ n ~ \textrm{is odd}, \\
                      \max\{\frac{1}{\sqrt{2}}I_k(\rho_3,\phi(0)),\frac{1}{\sqrt{2}}I_k(\rho_3,\phi(1))\}, & n=3.
                     \end{array}
 \right.
\end{equation}

The lower bound 1 in \cite{ChenMa2012} gives
\begin{equation}\label{eg1.Bound1GME}
C_{\textrm{GME}}\geq \left\{ \begin{array}{ll}
                     \max\{\frac{1}{\sqrt{2}(n-1)}I_k(\rho_n,\phi(0)),\frac{1}{\sqrt{2}(n-1)}I_k(\rho_n,\phi(1))\},  & n>3, \\
                      \max\{\frac{1}{2\sqrt{2}}I_k(\rho_3,\phi(0)),\frac{1}{2\sqrt{2}}I_k(\rho_3,\phi(1))\}, & n=3.
                     \end{array}
 \right.
\end{equation}

Obviously, for genuine multipartite entanglement measure,  our lower bound 1 Ineq.(\ref{eg1.Bound1k=2}) is better than that Ineq.(\ref{eg1.Bound1GME}) in \cite{ChenMa2012}.

The detection
parameter spaces of our bound 1 and bound 1 in \cite{ChenMa2012} of  genuine five-partite entanglement  are illustrated in Fig. 2 for the family $\rho_5$  of five-qubit states.  The area detected by our bound 1 is larger than the bound 1 of \cite{MaChen2011} when the two lower bounds are equal.

\begin{figure}
    \begin{center}
    {\includegraphics[scale=0.6]{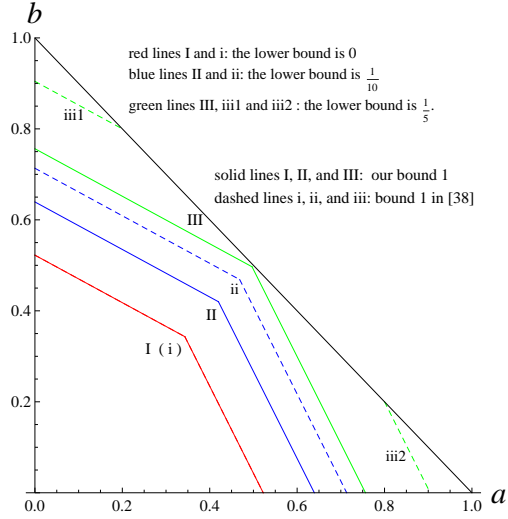}} \caption[Illustration]{(Color
online). The detection quality of our lower bound 1 and that in \cite{ChenMa2012}  on the genuine multipartite entanglement concurrence  is shown for the family $\rho_5=\frac{1-a-b}{32}I_{32}+a|W_5\rangle\langle W_5|+b|\tilde{W_5}\rangle\langle \tilde{W_5}|$  of five-qubit states, where $|W_5\rangle=\frac{1}{\sqrt{5}}(|00001\rangle+|00010\rangle+|00100\rangle+|01000\rangle+|10000\rangle)$ and $|\tilde{W_5}\rangle=\frac{1}{\sqrt{5}}(|11110\rangle+|11101\rangle+|11011\rangle+|10111\rangle+|01111\rangle)$.
The region above the line I (red) correspond to the genuine 5-partite entanglement detected by our bound 1, our criteria in \cite{GaoHong2010,GaoHong2012}, and the bound 1 of \cite{ChenMa2012}. The regions above the line II (blue) and the line III (green) correspond to the genuine 5-partite entanglement detected by our bound 1 when it is equal to or greater than $\frac{1}{5}$ and $\frac{1}{10}$, respectively.
 The states above the dashed line ii (blue),  the dashed line iii1 (green) and the dashed line iii2 (green), are detected by  the bound 1 of Ref.\cite{ChenMa2012} when it is equal to or greater than $\frac{1}{10}$, $\frac{1}{5}$ and $\frac{1}{5}$, respectively. Thus, the area detected by our bound 1 is visibly larger than that of \cite{ChenMa2012} when the two bounds are equal.  }
    \end{center}
    \end{figure}

Our Bound 2 Ineq.(\ref{bound-2}) is as follows:

\begin{equation}\label{eg1.bound2}
C_{k-ME}\geq \left\{ \begin{array}{ll}
                      \frac{1}{\sqrt{2}}H_k(I_k(\rho_n,\phi(0))+I_k(\rho_n,\phi(1))), & n>3, \\
                      \frac{1}{\sqrt{2}}H_k(I_k(\rho_3,\phi(0))+I_k(\rho_3,\phi(1))), & n=3.
                     \end{array}
 \right.
\end{equation}
Specially,
\begin{equation}\label{eg1.bound2k=2}
C_{2-ME}\geq \left\{ \begin{array}{ll}
                      \frac{\sqrt{2}}{n}(I_k(\rho_n,\phi(0))+I_k(\rho_n,\phi(1))),  & n>3 ~ \textrm{and} ~ n ~ \textrm{is even}, \\
                      \frac{\sqrt{2}}{\sqrt{n^2-1}}(I_k(\rho_n,\phi(0))+I_k(\rho_n,\phi(1))),  & n>3 ~ \textrm{and} ~ n ~ \textrm{is odd}, \\
                      \frac{1}{2}(I_k(\rho_3,\phi(0))+I_k(\rho_3,\phi(1))), & n=3.
                     \end{array}
 \right.
\end{equation}

The bound of Ref.\cite{MaChen2011} can not detect entanglement at all.
When $n\geq 4$, the lower bound 2 in \cite{ChenMa2012} can not detect entanglement at all.

Therefore, for the family of $n$-qubit states, the mixture of $W$ state and anti-$W$ state, dampened with white noise,  our lower bounds are better than the bounds 1 and  2 of Ref.\cite{ChenMa2012} and the bound of Ref.\cite{MaChen2011}.

\textit{Example 2}. Let us consider the family of $n$-qubit states
\begin{equation}\label{}
 \rho^{(G_n-W_n)}=\alpha|G_n\rangle\langle G_n|+\beta|W_n\rangle\langle W_n|+\frac{1-\alpha-\beta}{2^n}\textrm{I},
\end{equation}
the mixture of the GHZ state, the W state and the white noise. Here
$|G_n\rangle=\frac{1}{\sqrt{2}}(|00\cdots0\rangle+|11\cdots1\rangle)$
and
$|W_n\rangle=\frac{1}{\sqrt{n}}(|00\cdots001\rangle+|00\cdots010\rangle+\cdots+|10\cdots000\rangle)$.

For the selection $|\phi(0)\rangle=\otimes_{i=1}^n|x_i\rangle=|0\rangle^{\otimes n}$ and $|x'_i\rangle=|1\rangle$,  our bound 1 gives
\begin{equation}\label{}
C_{k-{\textrm{ME}}}(\rho^{(G_n-W_n)})\geq H_k [(n-1)\beta-n(n-1)\sqrt{(\frac{\alpha}{2}+\frac{1-\alpha-\beta}{2^n})\frac{1-\alpha-\beta}{2^n}}-n(n-k)(\frac{\beta}{n}+\frac{1-\alpha-\beta}{2^n})]. \end{equation}
Let $|\phi(x)\rangle=\otimes_{i=1}^n|x_i\rangle=\frac{|0\rangle-|1\rangle}{\sqrt{2}}^{\otimes n}$  and $|x'_i\rangle=\frac{|0\rangle+|1\rangle}{\sqrt{2}}$, our bound 1 gives
\begin{equation}\label{}
C_{k-{\textrm{ME}}}(\rho^{(G_n-W_n)})\geq
\begin{cases}
H_k[\frac{(n-1)(n-2)^2\beta}{2^n}-n(n-1)\sqrt{(\frac{1+\alpha-\beta}{2^n}+\frac{(n-4)^2\beta}{2^nn})\frac{1+\alpha-\beta+n\beta}{2^n}}-(n-k)(\frac{(n-2)^2\beta+n(1-\alpha-\beta)}{2^n})],\\
~~~~~~~~~~~~~~~~~~~~~~~~~~~~~~~~~~~~~~~~~~~~~~~~~~~~~~~~~~~~~~~~~~~~~~~~~~~~~~~~~~~~~~~~~~~~~~~~~~~~~~~~~n \textrm{~is~even},\\
\\
H_k[(n-1)\left(\frac{2n\alpha+(n-2)^2\beta}{2^n}-n\sqrt{(\frac{1-\alpha-\beta}{2^n}+\frac{(n-4)^2\beta}{2^nn})\frac{1-\alpha-\beta+n\beta}{2^n}}\right)-(n-k)(\frac{(n-2)^2\beta+n(1+\alpha-\beta)}{2^n})],\\
~~~~~~~~~~~~~~~~~~~~~~~~~~~~~~~~~~~~~~~~~~~~~~~~~~~~~~~~~~~~~~~~~~~~~~~~~~~~~~~~~~~~~~~~~~~~~~~~~~~~~~~~~n\textrm{~is~odd}.
\end{cases}
\end{equation}
For the selection $|\Phi\rangle=|0\rangle^{\otimes n}|1\rangle^{\otimes n}$, from (17) in \cite{MaChen2011}, there is
\begin{equation}\label{}
C_{\textrm{GME}}(\rho^{(G_n-W_n)})\geq
\begin{cases}
2[\frac{\alpha}{2}-C_n^1(\frac{\beta}{n}+\frac{1-\alpha-\beta}{2^n})^{\frac{1}{2}}(\frac{1-\alpha-\beta}{2^n})^{\frac{1}{2}}-(C_n^2+\cdots+\frac{1}{2}C_n^{\frac{n}{2}})(\frac{1-\alpha-\beta}{2^n})],&n \textrm{~is~even},\\
2[\frac{\alpha}{2}-C_n^1(\frac{\beta}{n}+\frac{1-\alpha-\beta}{2^n})^{\frac{1}{2}}(\frac{1-\alpha-\beta}{2^n})^{\frac{1}{2}}-(C_n^2+\cdots+C_n^{\lfloor\frac{n}{2}\rfloor})(\frac{1-\alpha-\beta}{2^n})],&n\textrm{~is~odd}.
\end{cases}
\end{equation}
Here $C_n^i$ is binomial coefficient, and $\lfloor\frac{n}{2}\rfloor$ is the nonnegative integer no more than  $\frac{n}{2}$.
Let $|\Phi\rangle=(\frac{|0\rangle+|1\rangle}{\sqrt{2}})^{\otimes n}(\frac{|0\rangle-|1\rangle}{\sqrt{2}})^{\otimes n}$, by (17) in \cite{MaChen2011}, there is
\begin{equation}\label{}
C_{\textrm{GME}}(\rho^{(G_5-W_5)})\geq2[\frac{15}{32}(1-\alpha+\frac{4\beta}{5})^{\frac{1}{4}}(1+\alpha+\frac{4\beta}{5})^{\frac{1}{4}}].
\end{equation}

The detection quality of our bound 1 and the bound in \cite{MaChen2011} on the genuine multipartite entanglement is illustrated in Fig. 3 for the family $ \rho^{(G_5-W_5)}$.
\begin{figure}
    \begin{center}
    {\includegraphics[scale=0.6]{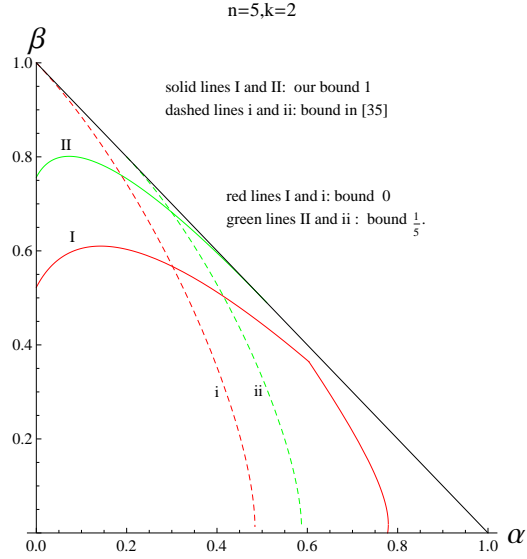}} \caption[Illustration]{(Color
online). The detection quality of our lower bound 1 and the bound in \cite{MaChen2011}  on the GME-concurrence  is shown for the family  of five-qubit states $\rho_5=\alpha|G_5\rangle\langle G_5|+\beta|W_5\rangle\langle W_5|+\frac{1-a-b}{32}I_{32}$ given by the convex combination of a GHZ state, a W state, and the maximally mixed state. The areas above the solid red line I and the dashed red line i are the genuine 5-partite entangled states  detected by our bound 1,  the bound  of \cite{MaChen2011}, respectively. The states in the areas above the solid green line II  (dashed green line ii) are genuine 5-partite entangled detected by our lower bound 1 (the bound  of \cite{MaChen2011})  when the  bound is equal to or greater than $\frac{1}{5}$.  }
    \end{center}
    \end{figure}

\section{experimental implementation of lower bounds}

The two lower bounds (\ref{bound-1}) and (\ref{bound-2})  are experimentally accessible by means of local observables, without quantum state tomography which requires an exponentially increasing measurements. Since nonlocal observable is not straightforward to measure in practice, the observables that can easily be measured in any experiment are local observables. In order to be useful in practice, measures for multipartite entanglement need to be experimentally implementable by means of local observables  without resorting to a full quantum state tomography. The lower bounds (\ref{bound-1}) and (\ref{bound-2}) satisfy these demands, as for fixed $|\phi(x)\rangle$, their computations only require at most $n^2+1$ and $2n^2+2$ measurements, respectively. Furthermore, they can be implemented locally as explicitly shown in \cite{GaoHong2012}. In total at most $\frac{5(n^2-n)}{2}+n+1$ and $5n^2-3n+2$ local observables are needed to implement our bound 1 and bound 2, respectively. In an experimental situation, it is now possible to choose the corresponding $|\phi(x)\rangle$ and not only detect the state as being $k$-nonseparable, but also have a reliable statement about the amount of multipartite entanglement the state exhibits.

\section{conclusion}

We have presented a measure of multipartite entanglement called $k$-ME concurrence that unambiguously detects all $k$-nonseparable states and studied multipartite entanglement of quantum states in arbitrary dimensional systems. This measure  satisfies important characteristics of an entanglement measure, such as entanglement monotone and vanishing on all $k$-separable states. Three main advantages are that $k$-ME concurrence is convex, subadditive and strictly greater than zero for all $k$-nonseparable states. The GME concurrence \cite{ChenMa2012,MaChen2011} is the special case of our $k$-ME concurrence when $k=2$.
Two powerful lower bounds of $k$-ME concurrence $C_{k-\mathrm{ME}}(\rho)$ for $n$-partite mixed quantum
states through the inequality (3) from Ref.\cite{GaoHong2012} are given.  We provide examples in which the lower bounds perform better than
the previously known methods.

\begin{acknowledgments}
This work was supported by the National Natural Science Foundation
of China under Grant No: 10971247, Hebei Natural Science Foundation
of China under Grant Nos: A2012205013, A2010000344, and  the Scientific Research Foundation for the Returned Overseas Chinese Scholars, Hebei Province.
\end{acknowledgments}

\appendix*

\section{Proof of two lower bounds}

Any pure quantum state of an
$n$ particle system can be denoted by vectors in Hilbert space
$\mathcal{H}=\mathcal{H}_{1}\otimes\mathcal{H}_{2}\otimes\cdots\otimes\mathcal{H}_{n}$,
as follows:
 \begin{equation}\label{}
|\psi\rangle=\sum\limits_{i_1,i_2,\cdots, i_n}c_{i_1i_2\cdots i_n}|i_1i_2\cdots i_n\rangle,
\end{equation}
which  can be rewritten as
\begin{equation}\label{}
|\psi\rangle=\sum\limits_{\gamma_{A_t}, \gamma_{\bar{A_t}}}c_{\gamma_{A_t}\gamma_{\bar{A_t}}}|\gamma_{A_t}\gamma_{\bar{A_t}}\rangle,
\end{equation}
where $\left\{ \left|i_{j}\right\rangle
\right\} $ is the orthonormal basis of $\mathcal{H}_{j}$, and
 a basis vector of subsystem $A_t$ is denoted by $|\gamma_{A_t}\rangle=|i_{j^t_1}i_{j^t_2}\cdots i_{j^t_{n_t}}\rangle$. Here    $A_1|A_2|\cdots|A_k=j^1_1 j^1_2\cdots j^1_{n_1}|j^2_1 j^2_2\cdots j^2_{n_2}|\cdots|j^k_1 j^k_2\cdots j^k_{n_k}$ is a $k$-partition of $\{1,2,\cdots,n\}$, and $\bar{A_t}$ is the complement of subsystem $A_t$ in $\{1,2,\cdots,n\}$. Thus,
\begin{equation}\label{}
\rho_{A_t}=\mathrm{Tr}_{\bar{A_t}}(|\psi\rangle\langle\psi|)
=\sum\limits_{\gamma_{A_t},\eta_{A_t}}(\sum\limits_{\gamma_{\bar{A_t}}}c_{\gamma_{A_t}\gamma_{\bar{A_t}}}c^*_{\eta_{A_t}\gamma_{\bar{A_t}}})|\gamma_{A_t}\rangle\langle\eta_{A_t}|
\equiv \sum\limits_{\gamma_{A_t},\eta_{A_t}}\rho_{\gamma_{A_t},\eta_{A_t}}|\gamma_{A_t}\rangle\langle\eta_{A_t}|,
\end{equation}
and
\begin{equation}\label{}
\begin{array}{rl}
\textrm{Tr}(\rho_{A_t}^2)=&\sum\limits_{\gamma_{A_t},\eta_{A_t}}|\rho_{\gamma_{A_t},\eta_{A_t}}|^2\\
=&\sum\limits_{\gamma_{A_t}}|\rho_{\gamma_{A_t},\gamma_{A_t}}|^2+2\sum\limits_{s_{\gamma_{A_t}}<s_{\eta_{A_t}}}|\rho_{\gamma_{A_t},\eta_{A_t}}|^2,
\end{array}
\end{equation}
where $s_{\gamma_{A_t}}=\sum_{l=1}^{n_t}i_{j^t_l} d_{j^t_l+1}d_{j^t_l+2}\cdots d_n d_{n+1}$ and $d_{n+1}=1$.  It follows that
\begin{equation}\label{1-tr}
\begin{array}{rl}
1-\textrm{Tr}(\rho_{A_t}^2)=&\sum\limits_{\gamma_{A_t}}\rho_{\gamma_{A_t},\gamma_{A_t}}(1-\rho_{\gamma_{A_t},\gamma_{A_t}})-2\sum\limits_{s_{\gamma_{A_t}}<s_{\eta_{A_t}}}|\rho_{\gamma_{A_t},\eta_{A_t}}|^2\\
=&2\sum\limits_{s_{\gamma_{A_t}}<s_{\eta_{A_t}}}(\rho_{\gamma_{A_t},\gamma_{A_t}}\rho_{\eta_{A_t},\eta_{A_t}}-|\rho_{\gamma_{A_t},\eta_{A_t}}|^2)\\
=&2\sum\limits_{s_{\gamma_{A_t}}<s_{\eta_{A_t}}}(\sum\limits_{\gamma_{\bar{A_t}},\eta_{\bar{A_t}}}|c_{\gamma_{A_t}\gamma_{\bar{A_t}}}c_{\eta_{A_t}\eta_{\bar{A_t}}}|^2
  -\sum\limits_{\gamma_{\bar{A_t}},\eta_{\bar{A_t}}}c_{\gamma_{A_t}\gamma_{\bar{A_t}}}c_{\eta_{A_t}\eta_{\bar{A_t}}}c^*_{\eta_{A_t}\gamma_{\bar{A_t}}}c^*_{\gamma_{A_t}\eta_{\bar{A_t}}})\\
=&2\sum\limits_{s_{\gamma_{A_t}}<s_{\eta_{A_t}}}\sum\limits_{s_{\gamma_{\bar{A_t}}}<s_{\eta_{\bar{A_t}}}}|c_{\gamma_{A_t}\gamma_{\bar{A_t}}}c_{\eta_{A_t}\eta_{\bar{A_t}}}-c_{\eta_{A_t}\gamma_{\bar{A_t}}}c_{\gamma_{A_t}\eta_{\bar{A_t}}}|^2.
\end{array}
\end{equation}

\subsection{Bound 1}

From (\ref{1-tr}) we have
\begin{equation}\label{k-concurrence-square}
\begin{array}{rl}
\frac{2\sum\limits_{t=1}^k(1-\textrm{Tr}(\rho^2_{A_t}))}{k}
=&\frac{4\sum\limits_{t=1}^k\sum\limits_{s_{\gamma_{A_t}}<s_{\eta_{A_t}}}\sum\limits_{s_{\gamma_{\bar{A_t}}}<s_{\eta_{\bar{A_t}}}}|c_{\gamma_{A_t}\gamma_{\bar{A_t}}}c_{\eta_{A_t}\eta_{\bar{A_t}}}-c_{\eta_{A_t}\gamma_{\bar{A_t}}}c_{\gamma_{A_t}\eta_{\bar{A_t}}}|^2}{k}\\
\geq&\frac{4\sum\limits_{t=1}^k\sum\limits_{|\eta_{A_t}|=1,|\eta_{\bar{A_t}}|=1}|c_{\eta_{A_t}0_{\bar{A_t}}}c_{0_{A_t}\eta_{\bar{A_t}}}-c_{0_{A_t}0_{\bar{A_t}}}c_{\eta_{A_t}\eta_{\bar{A_t}}}|^2}{k},
\end{array}
\end{equation}
where $0_{A_t}=(i_{j^t_1},i_{j^t_2},\cdots,i_{j^t_{n_t}})=(0,0,\cdots,0)$, $|\eta_{A_t}|$ and $|\eta_{\bar{A_t}}|$ represent the numbers of 1 in $\eta_{A_t}$, $\eta_{\bar{A_t}}$, respectively.

Next we deal with (\ref{k-concurrence-square}). By using the inequality $n\sum^n\limits_{i=1}|a_i|^2\geq(\sum^n\limits_{i=1}|a_i|)^2$ ($a_i$ is a complex number) and the triangle inequality, we obtain
\begin{equation}\label{}
\begin{array}{rl}
\sqrt{\frac{2\sum\limits_{t=1}^k(1-\textrm{Tr}(\rho^2_{A_t}))}{k}}
\geq&\frac{2}{\sqrt{k\sum\limits_{t=1}^kn_t(n-n_t)}}\sum\limits_{t=1}^k\sum\limits_{\substack{|\eta_{A_t}|=1\\|\eta_{\bar{A_t}}|=1}}(|c_{\eta_{A_t}0_{\bar{A_t}}}c_{0_{A_t}\eta_{\bar{A_t}}}-c_{0_{A_t}0_{\bar{A_t}}}c_{\eta_{A_t}\eta_{\bar{A_t}}}|)\\
\geq&\frac{2}{\sqrt{k\sum\limits_{t=1}^kn_t(n-n_t)}}\sum\limits_{t=1}^k\sum\limits_{\substack{|\eta_{A_t}|=1\\|\eta_{\bar{A_t}}|=1}}(|c_{\eta_{A_t}0_{\bar{A_t}}}c_{0_{A_t}\eta_{\bar{A_t}}}|-|c_{0_{A_t}0_{\bar{A_t}}}c_{\eta_{A_t}\eta_{\bar{A_t}}}|)\\
\geq & H_{k} Q_k,
\end{array}
\end{equation}
from which it follows
\begin{equation}\label{kCpure}
C_{k-\mathrm{ME}}(|\psi\rangle)=\min\limits_{A}\sqrt{\frac{2\sum\limits_{t=1}^k(1-\textrm{Tr}(\rho^2_{A_t}))}{k}}
\geq H_{k} Q_k,
\end{equation}
where
\begin{equation}\label{}
H_{k}=\min\limits_{A}\frac{\sqrt{k}}{\sqrt{\sum\limits_{t=1}^kn_t(n-n_t)}},
\end{equation}
 and
\begin{equation}\label{}
\begin{array}{rl}
Q_k=2\sum\limits_{\substack{s_{i_1\cdots i_n}<s_{l_1\cdots l_n}
\\|(i_1,\cdots, i_n)|=1\\|(l_1,\cdots, l_n)|=1}}|c_{i_1\cdots i_n}c_{l_1\cdots l_n}|-2\sum\limits_{|(i_1,\cdots, i_n)|=2}|c_{0\cdots0}c_{i_1\cdots i_n}|-(n-k)\sum\limits_{|(i_1,\cdots, i_n)|=1}|c_{i_1\cdots i_n}|^2.
\end{array}
\end{equation}
Here $|(i_1,\cdots, i_n)|$ denote the number of $i_l=1$ in $\{i_1,\cdots, i_n\}$.

Now suppose that $\rho=\sum\limits_mp_m\rho^m=\sum\limits_mp_m|\psi_m\rangle\langle\psi_m|$ is an $n$-partite mixed state where $|\psi_m\rangle=\sum\limits_{i_1,\cdots, i_n}c^m_{i_1\cdots i_n}|i_1\cdots i_n\rangle$. Using (\ref{kmix}) and (\ref{kCpure}), we see
\begin{equation}\label{kmixC}
C_{k-\mathrm{ME}}(\rho)=\inf\limits_{\{ p_m,|\psi_m\rangle\}}\sum\limits_m p_m C_{k-\mathrm{ME}}(|\psi_m\rangle)\geq H_k\inf\limits_{\{ p_m,|\psi_m\rangle\}}\sum\limits_mp_mQ_k^m.
\end{equation}
Let $|\phi(0)\rangle=|00\cdots 0\rangle$ and $0'=1$, we have
\begin{equation}\label{I_k}
\begin{array}{ll}
I_k(\rho,\phi(0))=   &       2\sum\limits_{i<j}|\rho_{\prod_{l=i+1}^{n+1}d_l,\prod_{l=j+1}^{n+1}d_l}|
-2\sum\limits_{i<j}\sqrt{\rho_{0,0}\rho_{\prod_{l=i+1}^{n+1}d_l+\prod_{l=j+1}^{n+1}d_l,\prod_{l=i+1}^{n+1}d_l+\prod_{l=j+1}^{n+1}d_l}} \\
      &    -(n-k)\sum\limits_i\rho_{\prod_{l=i+1}^{n+1}d_l,\prod_{l=i+1}^{n+1}d_l}.
\end{array}
\end{equation}
Here $d_{n+1}=1$.
Considering the three terms of (\ref{I_k}), we get
\begin{equation}\label{I_k1}
\begin{array}{rl}
2\sum\limits_{i<j}|\rho_{\prod_{l=i+1}^{n+1}d_l,\prod_{l=j+1}^{n+1}d_l}|\leq & 2 \sum\limits_{m}p_m\sum\limits_{i<j}|\rho^m_{\prod_{l=i+1}^{n+1}d_l,\prod_{l=j+1}^{n+1}d_l}|\\
=&\sum\limits_{m}p_m(2\sum\limits_{\substack{s_{i_1\cdots i_n}<s_{l_1\cdots l_n}
\\|(i_1, \cdots, i_n)|=1\\|(l_1, \cdots, l_n)|=1}}|c^m_{i_1\cdots i_n}c^m_{l_1\cdots l_n}|),
\end{array}
\end{equation}
\begin{equation}\label{I_k2}
\begin{array}{rl}
 & 2\sum\limits_{i<j}\sqrt{\rho_{0,0}\rho_{\prod_{l=i+1}^{n+1}d_l+\prod_{l=j+1}^{n+1}d_l,\prod_{l=i+1}^{n+1}d_l+\prod_{l=j+1}^{n+1}d_l}}
\\ =&2\sum\limits_{i<j}\sqrt{(\sum\limits_mp_m\rho^m_{0,0})(\sum\limits_mp_m\rho^m_{\prod_{l=i+1}^{n+1}d_l+\prod_{l=j+1}^{n+1}d_l,\prod_{l=i+1}^{n+1}d_l+\prod_{l=j+1}^{n+1}d_l})}\\
\geq & \sum\limits_mp_m(2\sum\limits_{|(i_1, \cdots, i_n)|=2}|c^m_{0\cdots0}c^m_{i_1\cdots i_n}|),
\end{array}
\end{equation}
\begin{equation}\label{I_k3}
(n-k)\sum\limits_i\rho_{\prod_{l=i+1}^{n+1}d_l,\prod_{l=i+1}^{n+1}d_l}=\sum\limits_mp_m(n-k)\sum\limits_{|(i_1, \cdots, i_n)|=1}|c^m_{i_1\cdots i_n}|^2.
\end{equation}
Combining (\ref{I_k1}), (\ref{I_k2}) and (\ref{I_k3}), we obtain
\begin{equation}\label{kmixI}
\begin{array}{rl}
I_k(\rho,\phi(0))\leq&\sum\limits_{m}p_m(2\sum\limits_{\substack{s_{i_1\cdots i_n}<s_{l_1\cdots l_n}
\\|(i_1, \cdots, i_n)|=1\\|(l_1, \cdots, l_n)|=1}}|c^m_{i_1\cdots i_n}c^m_{l_1\cdots l_n}|-2\sum\limits_{|(i_1, \cdots, i_n)|=2}|c^m_{0\cdots0}c^m_{i_1\cdots i_n}|\\
&-(n-k)\sum\limits_{|(i_1, \cdots, i_n)|=1}|c^m_{i_1\cdots i_n}|^2)\\
=&\sum\limits_{m}p_mQ_k^m,
\end{array}
\end{equation}
which implies that
\begin{equation}\label{}
I_k(\rho,\phi(0))\leq\inf\limits_{\{ p_m,|\psi_m\rangle\}}\sum\limits_mp_mQ_k^m.
\end{equation}
Therefore, from (\ref{kmixC}), there is
\begin{equation}\label{}
C_{k-\mathrm{ME}}(\rho)\geq H_kI_k(\rho,\phi(0)).
\end{equation}
Since for any fully separable state $|\phi(x)\rangle=\otimes_{i=1}^n|x_i\rangle=|x_1x_2\cdots x_n\rangle$, there exists a local unitary transformation $U=U_1\otimes U_2\otimes\cdots\otimes U_n$ such that $U|\phi(0)\rangle=|\phi(x)\rangle$, thus
 $H_kI_k(\rho,\phi(x))$ is also a lower bound because of the invariance of $C_{k-\mathrm{ME}}(\rho)$ under local unitary transformations. Therefore we have
\begin{equation}\label{bound-1-proof}
C_{k-\mathrm{ME}}(\rho)\geq\max\limits_{\{|\phi(x)\rangle\}}H_kI_k(\rho,\phi(x))\geq H_kI_k(\rho,\phi(x)),
\end{equation}
as desired.

Specially, when $k=2$, there is

\begin{equation}\label{bound1k=2}
 C_{2-\mathrm{ME}}(\rho)\geq \left\{ \begin{array}{ll}
                               \frac{2}{n}I_2(\rho,\phi(x)), & n ~ \textrm{is even}, \\
                               \frac{2}{\sqrt{n^2-1}}I_2(\rho,\phi(x)), & n ~ \textrm{is odd}.
                             \end{array} \right.
\end{equation}

Our lower bound 1 (\ref{bound1k=2}) is greater than $\frac{1}{\sqrt{2}(n-1)}$, i.e. our lower bound 1 is stronger than that in \cite{ChenMa2012}.

\subsection{Bound 2}

By (\ref{1-tr}), we get
\begin{equation}\label{}
\begin{array}{rl}
\frac{2\sum\limits_{t=1}^k(1-\textrm{Tr}(\rho^2_{A_t}))}{k}=&\frac{4\sum\limits_{t=1}^k\sum\limits_{s_{\gamma_{A_t}}<s_{\eta_{A_t}}}
\sum\limits_{s_{\gamma_{\bar{A_t}}}<s_{\eta_{\bar{A_t}}}}|c_{\gamma_{A_t}\gamma_{\bar{A_t}}}c_{\eta_{A_t}\eta_{\bar{A_t}}}-c_{\eta_{A_t}\gamma_{\bar{A_t}}}c_{\gamma_{A_t}\eta_{\bar{A_t}}}|^2}{k}\\
\geq&\frac{4\sum\limits_{t=1}^k(\sum\limits_{\substack{|\eta_{A_t}|=1 \\ |\eta_{\bar{A_t}}|=1}}|c_{\eta_{A_t}0_{\bar{A_t}}}c_{0_{A_t}\eta_{\bar{A_t}}}-c_{0_{A_t}0_{\bar{A_t}}}c_{\eta_{A_t}\eta_{\bar{A_t}}}|^2
+\sum\limits_{\substack{|\eta_{A_t}|=n_t-1 \\ |\eta_{\bar{A_t}}|=n-n_t-1}}|c_{\eta_{A_t}1_{\bar{A_t}}}c_{1_{A_t}\eta_{\bar{A_t}}}-c_{1_{A_t}1_{\bar{A_t}}}c_{\eta_{A_t}\eta_{\bar{A_t}}}|^2)}{k}.
\end{array}
\end{equation}
Similar to the proof of bound 1, there is
\begin{equation}\label{}
\begin{array}{rl}
\sqrt{\frac{2\sum\limits_{t=1}^k(1-\textrm{Tr}(\rho^2_{A_t}))}{k}}
\geq&\frac{2}{\sqrt{2k\sum\limits_{t=1}^kn_t(n-n_t)}}\sum\limits_{t=1}^k(\sum\limits_{\substack{|\eta_{A_t}|=1\\|\eta_{\bar{A_t}}|=1}}
|c_{\eta_{A_t}0_{\bar{A_t}}}c_{0_{A_t}\eta_{\bar{A_t}}}-c_{0_{A_t}0_{\bar{A_t}}}c_{\eta_{A_t}\eta_{\bar{A_t}}}|
\\ & +\sum\limits_{\substack{|\eta_{A_t}|=n_t-1\\|\eta_{\bar{A_t}}|=n-n_t-1}}|c_{\eta_{A_t}1_{\bar{A_t}}}c_{1_{A_t}\eta_{\bar{A_t}}}
-c_{1_{A_t}1_{\bar{A_t}}}c_{\eta_{A_t}\eta_{\bar{A_t}}}|)\\
\geq&\frac{2}{\sqrt{2k\sum\limits_{t=1}^kn_t(n-n_t)}}\sum\limits_{t=1}^k[\sum\limits_{\substack{|\eta_{A_t}|=1\\|\eta_{\bar{A_t}}|=1}}
(|c_{\eta_{A_t}0_{\bar{A_t}}}c_{0_{A_t}\eta_{\bar{A_t}}}|-|c_{0_{A_t}0_{\bar{A_t}}}c_{\eta_{A_t}\eta_{\bar{A_t}}}|) \\
& +\sum\limits_{\substack{|\eta_{A_t}|=n_t-1\\|\eta_{\bar{A_t}}|=n-n_t-1}}(|c_{\eta_{A_t}1_{\bar{A_t}}}c_{1_{A_t}\eta_{\bar{A_t}}}|-|c_{1_{A_t}1_{\bar{A_t}}}c_{\eta_{A_t}\eta_{\bar{A_t}}}|)]\\
\geq & \frac{\sqrt{k}}{\sqrt{2\sum\limits_{t=1}^kn_t(n-n_t)}}(Q_k+\bar{Q}_k).
\end{array}
\end{equation}
So, we get
\begin{equation}\label{kCpure-2}
\begin{array}{rl}
C_{k-\mathrm{ME}}(|\psi\rangle)
\geq &\bar{H}_{k}(Q_k+\bar{Q}_k),
\end{array}
\end{equation}
where
\begin{equation}\label{}
\bar{H}_{k}=\min\limits_{A}\frac{\sqrt{k}}{\sqrt{2\sum\limits_{t=1}^kn_t(n-n_t)}}=\frac{H_k}{\sqrt{2}},
\end{equation}
\begin{equation}\label{}
\begin{array}{rl}
Q_k=&2\sum\limits_{\substack{s_{i_1\cdots i_n}<s_{l_1\cdots l_n}
\\|(i_1, \cdots, i_n)|=1\\|(l_1, \cdots, l_n)|=1}}|c_{i_1\cdots i_n}c_{l_1\cdots l_n}|-2\sum\limits_{|(i_1, \cdots, i_n)|=2}|c_{0\cdots0}c_{i_1\cdots i_n}|-(n-k)\sum\limits_{|(i_1, \cdots, i_n)|=1}|c_{i_1\cdots i_n}|^2.
\end{array}
\end{equation}
\begin{equation}\label{}
\begin{array}{cc}
\bar{Q}_k=&2\sum\limits_{\substack{s_{i_1\cdots i_n}<s_{l_1\cdots l_n}
\\|(i_1, \cdots, i_n)|=n-1\\|(l_1, \cdots, l_n)|=n-1}}|c_{i_1\cdots i_n}c_{l_1\cdots l_n}|-2\sum\limits_{|(i_1, \cdots, i_n)|=n-2}|c_{1\cdots1}c_{i_1\cdots i_n}|-(n-k)\sum\limits_{|(i_1, \cdots, i_n)|=n-1}|c_{i_1\cdots i_n}|^2.
\end{array}
\end{equation}

Now suppose that $\rho=\sum\limits_mp_m\rho^m=\sum\limits_mp_m|\psi_m\rangle\langle\psi_m|$ is an $n$-partite mixed state where $|\psi_m\rangle=\sum\limits_{i_1,\cdots, i_n}c^m_{i_1\cdots i_n}|i_1\cdots i_n\rangle$. Using (\ref{kmix}) and (\ref{kCpure-2}), we see
\begin{equation}\label{kmixC-2}
C_{k-\mathrm{ME}}(\rho)=\inf\limits_{\{ p_m,|\psi_m\rangle\}}\sum\limits_m p_m C_{k-\mathrm{ME}}(|\psi_m\rangle)\geq \bar{H}_k\inf\limits_{\{ p_m,|\psi_m\rangle\}}\sum\limits_mp_m(Q_k^m+\bar{Q}_k^m).
\end{equation}

Let $|\phi(1)\rangle=|11\cdots 1\rangle$ and $1'=0$, then there is
\begin{equation}\label{I_k'}\begin{array}{ll}
 I_k(\rho,\phi(1))= & 2\sum\limits_{i<j}|\rho_{\sum\limits_{l\neq i} d_{l+1}d_{l+2}\cdots d_{n+1},\sum\limits_{l\neq j} d_{l+1}d_{l+2}\cdots d_{n+1}}| \\
 & -2\sum\limits_{i<j}\sqrt{\rho_{\sum\limits_{l} d_{l+1}d_{l+2}\cdots d_{n+1},\sum\limits_{l} d_{l+1}d_{l+2}\cdots d_{n+1}}\rho_{\sum\limits_{l\neq i,j} d_{l+1}d_{l+2}\cdots d_{n+1},\sum\limits_{l\neq i,j} d_{l+1}d_{l+2}\cdots d_{n+1}}} \\
 & -(n-k)\sum\limits_i\rho_{\sum\limits_{l\neq i} d_{l+1}d_{l+2}\cdots d_{n+1},\sum\limits_{l\neq i} d_{l+1}d_{l+2}\cdots d_{n+1}},
 \end{array}
\end{equation}
where $d_{n+1}=1$.
For the first term of (\ref{I_k'}),
\begin{equation}\label{I_k1'}
\begin{array}{rl}
2\sum\limits_{i<j}|\rho_{\sum\limits_{l\neq i} d_{l+1}d_{l+2}\cdots d_{n+1},\sum\limits_{l\neq j} d_{l+1}d_{l+2}\cdots d_{n+1}}|\leq & \sum\limits_{m}p_m(2\sum\limits_{\substack{s_{i_1\cdots i_n}<s_{l_1\cdots l_n}
\\|(i_1, \cdots, i_n)|=n-1\\|(l_1, \cdots, l_n)|=n-1}}|c^m_{i_1\cdots i_n}c^m_{l_1\cdots l_n}|).
\end{array}
\end{equation}
For the second term,
\begin{equation}\label{I_k2'}
\begin{array}{rl}
& 2\sum\limits_{i<j}\sqrt{\rho_{\sum\limits_{l} d_{l+1}d_{l+2}\cdots d_{n+1},\sum\limits_{l} d_{l+1}d_{l+2}\cdots d_{n+1}}\rho_{\sum\limits_{l\neq i,j} d_{l+1}d_{l+2}\cdots d_{n+1},\sum\limits_{l\neq i,j} d_{l+1}d_{l+2}\cdots d_{n+1}}}\\
=& 2\sum\limits_{i<j}\sqrt{(\sum\limits_mp_m\rho^m_{\sum\limits_{l} d_{l+1}d_{l+2}\cdots d_{n+1},\sum\limits_{l} d_{l+1}d_{l+2}\cdots d_{n+1}})(\sum\limits_mp_m\rho^m_{\sum\limits_{l\neq i,j} d_{l+1}d_{l+2}\cdots d_{n+1},\sum\limits_{l\neq i,j} d_{l+1}d_{l+2}\cdots d_{n+1}})}\\
\geq &\sum\limits_mp_m (2\sum\limits_{|(i_1, \cdots, i_n)|=n-2}|c^m_{1\cdots1}c^m_{i_1\cdots i_n}|).
\end{array}
\end{equation}
For the third term,
\begin{equation}\label{I_k3'}
(n-k)\sum\limits_i\rho_{\sum\limits_{l\neq i} d_{l+1}d_{l+2}\cdots d_{n+1},\sum\limits_{l\neq i} d_{l+1}d_{l+2}\cdots d_{n+1}}=\sum\limits_mp_m[(n-k)\sum\limits_{|(i_1, \cdots, i_n)|=n-1}|c_{i_1\cdots i_n}|^2].
\end{equation}
Combining (\ref{I_k1'}), (\ref{I_k2'}) and (\ref{I_k3'}) gives that
\begin{equation}\label{kmixI'}
\begin{array}{rl}
I_k(\rho,\phi(1))\leq &\sum\limits_{m}p_m\bar{Q}_k^m.
\end{array}
\end{equation}

From (\ref{kmixC-2}), (\ref{kmixI}) and (\ref{kmixI'}), we obtain
\begin{equation}\label{}
C_{k-\mathrm{ME}}(\rho)\geq \bar{H}_k(I_k(\rho,\phi(0))+I_k(\rho,\phi(1))).
\end{equation}
Note that for any fully separable state $|\phi(x)\rangle=\otimes_{i=1}^n |x_i\rangle$, there is a local unitary transformation $V=V_1\otimes V_2\otimes\cdots \otimes V_n$ satisfying  $V|\phi(0)\rangle=|\phi(x)\rangle$ and $V|\phi(1)\rangle=|\phi(y)\rangle$. Thus   $\bar{H}_k(I_k(\rho,\phi(x))+I_k(\rho,\phi(y)))$ is also a lower bound because of the invariance of $C_{k}(\rho)$ under local unitary transformations, so we have
\begin{equation}\label{bound-2-proof}
C_{k-\mathrm{ME}}(\rho)\geq\max\limits_{\{\phi(x),\phi(y)\}}\bar{H}_k(I_k(\rho,\phi(x))+I_k(\rho,\phi(y)))\geq \bar{H}_k(I_k(\rho,\phi(x))+I_k(\rho,\phi(y))).
\end{equation}
Here $|\phi(x)\rangle=\otimes_{i=1}^n|x_i\rangle$ and $|\phi(y)\rangle=\otimes_{i=1}^n|y_i\rangle$ are orthogonal full separable states.
The proof is complete.

\end{document}